# Multi-Scale Times and Modes of Fast and Slow Relaxation in Solutions with Coexisting Spherical and Cylindrical Micelles according to the Difference Becker-Döring Kinetic Equations

Ilya A. Babintsev, Loran Ts. Adzhemyan, Alexander K. Shchekin*

*Department of Statistical Physics, Faculty of Physics, St Petersburg State University, Ulyanovskaya 1, Petrodvoretz, St Petersburg, 198504, Russian Federation*

The eigenvalues and eigenvectors of the matrix of coefficients of the linearized kinetic equations applied to aggregation in surfactant solution determine the full spectrum of characteristic times and specific modes of micellar relaxation. The dependence of these relaxation times and modes on the total surfactant concentration has been analyzed for concentrations in the vicinity and well above the second critical micelle concentration ($cmc_2$) for systems with coexisting spherical and cylindrical micelles. The analysis has been done on the basis of a discrete form of the Becker-Döring kinetic equations employing the Smoluchowsky diffusion model for the attachment rates of surfactant monomers to surfactant aggregates with matching the rates for spherical aggregates and the rates for large cylindrical micelles. The equilibrium distribution of surfactant aggregates in solution has been modeled as having one maximum for monomers, another maximum for spherical micelles and wide slowly descending branch for cylindrical micelles. The results of computations have been compared with the analytical ones known in the limiting cases from solutions of the continuous Becker-Döring kinetic equation. They demonstrated a fair agreement even in the vicinity of the $cmc_2$ where the analytical theory looses formally its applicability.

## I. INTRODUCTION

Micellization, i.e. spontaneous formation of a large number of stable molecular aggregates or micelles, can be observed in surfactant solutions with chained surfactant molecules after increasing the total surfactant concentration to or above the critical micelle concentration (cmc). It is a typical barrier process of self-assembly of surfactant molecules into aggregates which tends to establishing aggregative equilibrium in solution between aggregates with different aggregation numbers.[1-6] The aggregates may have different shapes; the spherical micelles appear starting from the first cmc ($cmc_1$), and cylindrical micelles aggregate can be observed above the second cmc ($cmc_2$). Disturbances of the aggregative equilibrium by temperature, pressure or concentration jumps relax through several time stages to a new aggregative equilibrium with redistribution of aggregates in the their size and shape.[7] Kinetics of micellization and relaxation in micellar solutions depends on structural peculiarities of molecular aggregates, their mobility in solutions, the character of micelle-monomer processes and micelle-micelle interactions.[8,9] Understanding these dependences opens new possibilities for control and design of micellar systems for various applications.

The foundations of the kinetic theory of relaxation in micellar solutions with spherical micelles were build by Aniansson and Wall,[10-12] Almgren et al.,[13] Teubner,[14] and Kahlweit.[15] A review on the subject had been given by Kahlweit and Teubner[16] and Zana.[7] The general approach assumes that kinetics of micellization and micellar relaxation is governed by the stepwise molecular mechanism via attachment-detachment of single surfactant molecules. This mechanism can be described on the basis of the Becker-Döring kinetic equations for evolution of aggregate concentrations with different aggregation numbers.[4,17-23] According to some experimental evidence,[24,25] another mechanism with fusion and fission of aggregates[23,26-29] is not important, at least, at not very high total surfactant concentrations. In last decade, the theoretical kinetic description in frameworks of the Becker-Döring equations were extended to solutions with coexisting spherical and cylindrical micelles.[30-35] It has been shown that polydispersity of cylindrical aggregates and transitions between spherical and cylindrical micelles introduce new features in the kinetic behaviour of micellar systems at small and large time scales. However, the results of analytical theory were limited by the region of total concentrations which far exceed the $cmc_2$.

Recently we presented a numerical study of micellization and relaxation to tintermediate quasi-equilibrium states and final complete equilibrium in surfactant solution with nonionic spherical micelles[36] and cylindrical micelles.[37] This study described overall time behavior of surfactant monomer and aggregate concentrations at micellization and relaxation under arbitrary initial deviations from equilibrium at all time scales. It has been developed on the basis of direct numerical computations with using a discrete form of the full and linearized Becker-Döring kinetic equations and the simplified droplet model for spherical aggregates[36,38-41] and combined droplet and linear model for cylindrical aggregates.[31,37,42-44] Such approach has several advantages over molecular dynamics simulations because it is free of limitations of the analytical theory, allows one to use reasonable models for the aggregation work and attachment coefficients and trace in detail the kinetic behavior of micellar systems for any aggregation numbers and total surfactant concentration.

The same approach is used also in this paper which presents the results of numerical analysis of relaxation in nonionic surfactant solution with coexisting spherical premicellar aggregates, stable spherical micelles and stable polydisperse cylindrical micelles for different total surfactant concentrations close to and exceeding the $cmc_2$. We focus here on finding the full spectrum of characteristic times and specific modes of micellar relaxation in their dependence on the total surfactant concentration. The numerical results will be compared with the analytical ones known in the limiting cases from solutions of the continuous Becker-Döring kinetic equation.[30-35]

The paper is organized as follows. The models for the equilibrium distribution of aggregates in aggregate numbers and the aggregate-monomer attachment coefficients are considered in Section II. The full discrete spectrum of characteristic times of micellar relaxation and relaxation modes computed through the analysis of eigenvalues and eigenvectors of the linearized matrix of the Becker-Döring difference equations are presented in Section III. Dependence of the characteristic times and specific modes of relaxation on the total surfactant concentration and comparison with predictions of analytical kinetic theory of relaxation for the linearized continuous Becker-Döring equation is given in Section IV. The results are summarised in Section V.

## II. MODELS FOR THE AGGREGATION WORK AND THE ATTACHMENT COEFFICIENTS

**Equilibrium distribution and work of aggregation.** Let us consider a surfactant solution with the total surfactant concentration being close to or above the cmc$_2$. We will be interested in description of non-equilibrium and equilibrium aggregation states in such a solution where premicellar spherical aggregates and spherical micelles, transient (in shape from spherical to cylindrical) aggregates, and large cylindrical micelles coexist and transform into each other. Because any surfactant molecule can serve as a nucleus for formation of an aggregate, the equilibrium concentration $c_n^{eq}$ of aggregates with aggregation number $n$ satisfies the Boltzmann distribution which is related to concentration $c_1$ of surfactant monomers ($n=1$) and the dimensionless aggregation work $W_n$ (expressed in thermal energy units $k_B T$ where $k_B$ is the Boltzmann constant and $T$ is the absolute temperature of solution) in the form

$$c_n^{eq} = c_1 e^{-W_n} . \qquad (1)$$

Because work $W_n$ is a function of aggregation number $n$ and monomer concentration $c_1$, it is convenient to introduce an independent of monomer concentration work $\bar{W}_n$. In the case of ideal mixture of aggregates,[4,20]

$$\bar{W}_n = W_n + (n-1)\ln c_1 . \qquad (2)$$

Here and below, the monomer concentration $c_1$ is assumed to be measured in units of concentration at which $W_n$ coincides with $\bar{W}_n$. With the help of work $\bar{W}_n$, eq.(1) can be rewritten as

$$c_n^{eq} = c_1^n e^{-\bar{W}_n} . \qquad (2a)$$

In the vicinity and above the cmc$_2$, the aggregation work $W_n$ should have two maxima at points $n_c^{(1)}$ and $n_c^{(2)}$, two minima at points $n_s^{(1)}$ and $n_s^{(2)}$ and slowly increasing linear tail at larger $n$.[30,31,33-35,42,43] Evidently, the values $n_c^{(1)}$, $n_c^{(2)}$, $n_s^{(1)}$, and $n_s^{(2)}$ depend on the surfactant monomer concentration. Assuming that concentration $c_1 = 1$ corresponds to the total concentration in the vicinity of cmc$_2$, we can propose the following model for the aggregation work $\bar{W}_n$:

$$\bar{W}_n = \begin{cases} w_1(n-1)^{4/3} + w_2(n-1) + w_3(n-1)^{2/3}, & 1 \leq n \leq \bar{n}_s^{(1)} \\ v_1(n-\bar{n}_s^{(1)})^4 + v_2(n-\bar{n}_s^{(1)})^3 + v_3(n-\bar{n}_s^{(1)})^2 + \bar{W}_s^{(1)}, & \bar{n}_s^{(1)} \leq n \leq \bar{n}_0 \\ \bar{k}(n-\bar{n}_0) + \bar{W}_0, & n > \bar{n}_0 \end{cases} \qquad (3)$$

where the values $\bar{n}_s^{(1)}$ and $\bar{n}_s^{(2)}$ have the same meaning as the values $n_s^{(1)}$ and $n_s^{(2)}$, but do not depend on concentration $c_1$, $\bar{W}_s^{(1)}$ is the value of the first minimum of the work $\bar{W}_n$, $\bar{W}_0 \equiv \bar{W}_n\big|_{n=\bar{n}_0}$, and $\bar{n}_0$ is the starting point for linear growth of work $\bar{W}_n$. It is assumed in eq.(3) that work $\bar{W}_n$ for aggregates with aggregation numbers $n \leq \bar{n}_s^{(1)}$ corresponds to the simplified droplet model for spherical aggregate[4,36,38-41] with maximum at $\bar{n}_c^{(2)}$ and minimum at $\bar{n}_s^{(1)}$, while the work at $n > \bar{n}_0$ refers to the linear model[26,30,31,33-35,37] for cylindrical aggregates. In the transient (from spherical to cylindrical aggregates) range $\bar{n}_s^{(1)} < n \leqslant \bar{n}_0$, we use a polynomial interpolation having a maximum at $\bar{n}_c^{(1)}$ and two minima at $\bar{n}_s^{(1)}$ and $\bar{n}_s^{(2)}$. As parameters of the work $\bar{W}_n$, we have fixed several characteristic points: the locations $\bar{n}_s^{(1)}$ and $\bar{n}_s^{(2)}$ of two minima of the work, the values $\bar{W}_s^{(1)}$ and $\bar{W}_s^{(2)}$ of these minima themselves and the values $\bar{W}_c^{(1)}$ and $\bar{W}_c^{(2)}$ ($\bar{W}_c^{(1)} > \bar{W}_c^{(2)}$) of the maxima of the work (but not the locations of these maxima). These six conditions determine six parameters $w_i$ and $v_i$ ($i=1,2,3$), while additional two conditions of continuity of function $\bar{W}_n$ and its derivative with respect to aggregation number at $n = \bar{n}_0$ at fixed $\bar{k}$ determine $\bar{n}_0$ and $\bar{W}_0$. It should be noted that the selection of the point of minimum $\bar{n}_1^{(s)}$ as a point of patching the droplet and transient models for aggregates is convenient for finding the parameters of the aggregation work, but has a major drawback because it not ensures equality of second derivatives of the model functions for $\bar{W}_n$ at the minimum and makes asymmetric potential well even for small deviations from the minimum point. However we will minimize the degree of this asymmetry by choosing appropriate values of parameters.

In order to extend the range of aggregation numbers and total concentrations for comparison with the results of the analytical theory, we will consider below two sets of parameters determining two works $\bar{W}_n$ (work 1 and work 2). Let us take for the work 1 the following set of fixed a priori parameters:

$$\bar{W}_c^{(1)} = 15 , \; \bar{n}_s^{(1)} = 100 , \; \bar{W}_s^{(1)} = 5 ,$$
$$\bar{W}_c^{(2)} = 14 , \; \bar{n}_s^{(2)} = 300 , \; \bar{W}_s^{(2)} = 9 , \; \bar{k} = 0.01 . \qquad (4)$$

As a consequence, the parameters of the work 1 entering Eq.(3) are

$$\bar{n}_c^{(1)} = 16 , \; \bar{n}_c^{(2)} = 211 , \; w_1 = 0.4317 , \; w_2 = -4.0955 ,$$
$$w_3 = 9.9403 , \; v_1 = 6.8358 \cdot 10^{-8} , \; v_2 = -2.8343 \cdot 10^{-5} ,$$
$$v_3 = 3.0343 \cdot 10^{-3} , \; \bar{n}_0 = 301 , \; \bar{W}_0 = 9.0025 . \qquad (5)$$

The fixed parameters of work 2 are taken as

$$\bar{W}_c^{(1)} = 20 , \; \bar{n}_s^{(1)} = 21 , \; \bar{W}_s^{(1)} = 7 ,$$
$$\bar{W}_c^{(2)} = 15 , \; \bar{n}_s^{(2)} = 50 , \; \bar{W}_s^{(2)} = 12 , \; \bar{k} = 0.02 . \qquad (6)$$

These parameters provide for work 2

$$\bar{n}_c^{(1)} = 4, \quad \bar{n}_c^{(2)} = 38, \quad w_1 = 4.8007, \quad w_2 = -26.762,$$
$$w_3 = 38.222, \quad v_1 = 1.1676 \cdot 10^{-4}, \quad v_2 = -7.182 \cdot 10^{-3},$$
$$v_3 = 0.11603, \quad \bar{n}_0 = 50, \quad \overline{W}_0 = 12. \quad (7)$$

The detailed behavior of work 1 and work 2 as functions of aggregation number $n$ is clearly seen in Fig.1. The asymmetry of each work in the vicinity of minimum at $n = \bar{n}_s^{(1)}$ is small. Note that work 1 and work 2 are determined in different total ranges of aggregation numbers. Maximal value $n_m$ of the aggregation number for work 1 equals $n_m^{(w1)} = 3500$, while the same number for work 2 equals $n_m^{(w2)} = 5000$.

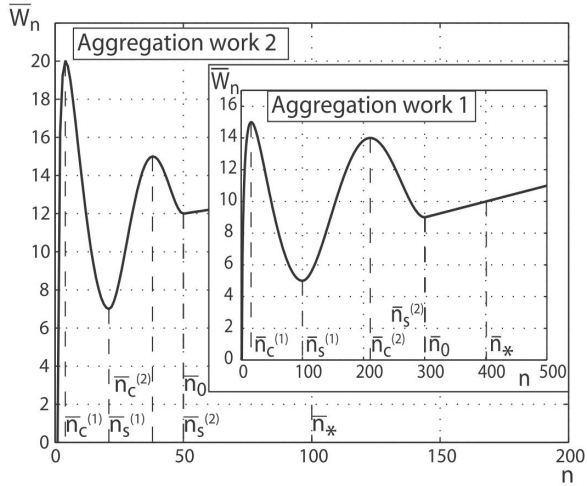

**Fig.1** The micellization work $\overline{W}_n$ as a function of the aggregation number $n$.

**The micellization degree.** The most important distinction between work 1 and work 2 is that they differ in prediction of the first and second critical micelle concentrations (cmc$_1$ and cmc$_2$). The cmc can be determined as the total surfactant concentration at which the equilibrium micellization degree equals 10%. In our notation, the micellization degree $\alpha_1^{eq}(c_1)$ for spherical micelles can be written as a function of monomer surfactant concentration as

$$\alpha_1^{eq}(c_1) \equiv \sum_{n=n_c^{(1)}}^{n_c^{(2)}} n c_n^{eq}(c_1) \bigg/ \sum_{n=1}^{n_m} n c_n^{eq}(c_1) . \quad (8)$$

In the same way, the micellization degree $\alpha_2^{eq}(c_1)$ for cylindrical micelles as a function of monomer surfactant concentration has the form

$$\alpha_2^{eq}(c_1) \equiv \sum_{n=n_c^{(2)}}^{n_m} n c_n^{eq}(c_1) \bigg/ \sum_{n=1}^{n_m} n c_n^{eq}(c_1) . \quad (9)$$

The dependence of the equilibrium micellisation degrees for spherical and cylindrical micelles on the monomer concentration $c_1$ computed with the help of eqs. (2)-(9) for two aggregation works is shown in Fig.2. As follows from Fig.2, the cmc$_1$ and cmc$_2$ for work 1 and for work 2 are reached at the monomer concentrations $(c_1)_{cmc_1}^{(1)} = 0.94385$, $(c_1)_{cmc_2}^{(1)} = 0.9977$ and $(c_1)_{cmc_1}^{(2)} = 0.9965$, $(c_1)_{cmc_2}^{(2)} = 1.0091$, correspondingly. We can conclude that contribution of spherical micelles to the total micellization degree at cmc$_2$ in the case of work 1 is eight times larger than corresponding contribution of cylindrical micelles while both contributions are almost the same in the case of work 2.

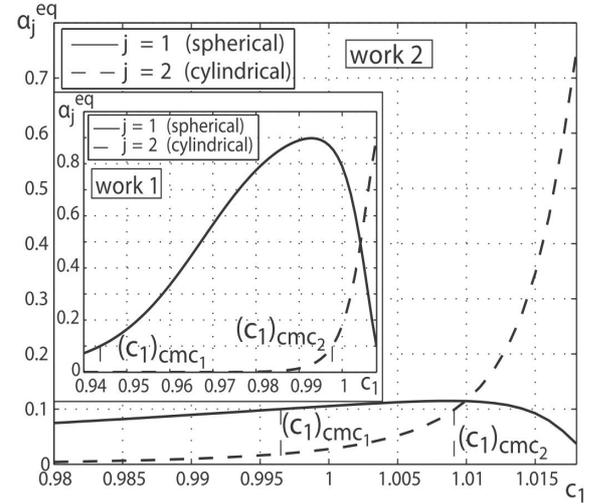

**Fig.2** The equilibrium micellization degree $\alpha^{eq}$ for spherical and cylindrical micelles as a function of monomer concentration $c_1$ for aggregation work 1 and work 2.

**Attachment coefficients.** The important kinetic characteristics of aggregation process in micellar system are the aggregate-monomer attachment coefficients $a_n$ determined in their dependence on the aggregation number $n$. The model for the aggregate-monomer attachment coefficients $a_n$ should correspond to that for the aggregation work $\overline{W}_n$ in eq. (3) and be different for spherical aggregates at aggregation numbers $1 \leq n \leq n_s^{(1)}$, for transient aggregates at aggregation numbers $n_s^{(1)} \leq n \leq \bar{n}_0$, and for cylindrical aggregates at $n > \bar{n}_0$. Considering stochastic movements of molecular aggregates in solution as obeying to brownian diffusion kinetics,[45,46] one can suggest that attachments of monomers to a spherical aggregate proceed with a stationary diffusion-controlled rate. When both the monomer and the aggregate move, then $a_n \propto (R_1 + R_n)(D_1 + D_n)$ where $R_n$ and $D_n$ represent the radius and diffusion coefficient of an aggregate $\{n\}$ in surfactant solution, including monomers as a particular case of aggregates with $n = 1$. Taking into account that the Stokes–Einstein formula for spherical aggregates provides $D_n \propto 1/R_n$ and the droplet model of a spherical micelle[4,39-41] assumes $R_n \propto n^{1/3}$ allow one to write at $1 \ll n < n_c^{(2)}$: $a_n \propto n^{1/3}$. At $n \geq \bar{n}_0$ the diffusion coefficient for cylindrical micelles becomes small, and it is sufficient for calculation of $a_n$ to find the stationary flux of monomers onto nonmovable cylindrical body in polar coordinates. This flux will be proportional the length of the cylindrical body, and, because the radius of the body is fixed for cylindrical micelles,[30,31,43,44] the length of the body itself is proportional to the aggregation number $n$. Thus we have $a_n \propto n$ at $n > \bar{n}_0$. In view of the above consideration, we will use the following continuous model for the attachment coefficients $a_n$ at arbitrary $n$:

$$a_n = \frac{n^{1/3}(n+\bar{n}_0)^{2/3}}{\bar{n}_0}, \quad 1 \le n \le n_m - 1, \quad a_{n_m} = 0. \quad (10)$$

Here we match the asymtotic cases for spherical aggregates at $n \ll \bar{n}_0$ and for cylindrical aggregates at $n \gg \bar{n}_0$ and include the specific factor appeared from the proportionality constants as a scale into the quantity $a_n$. As is clear, the attachment coefficient $a_n$ should have physically a dimensionality of reciprocal time. Representation the quantity $a_n$ in the form (10) means that below we will consider a dimensionless time.

## III. THE CHARACTERISTIC TIMES AND SPECIFIC MODES OF RELAXATION

**Kinetic equation of aggregation.** Kinetics of stepwise formation and fragmentation of aggregates with diffrent aggregation numbers under nonequilibrium conditions in surfactant micellar solutions is governed by the system of the Becker-Döring difference equations for non-equilibrium aggregate concentrations $c_n(t)$ as functions of time $t$.[4,11-16,20,21] For an isolated system with fixed total surfactant concentration $C = \sum_{n=1}^{n_m} n c_n$ and finite upper aggregation number $n_m$, the Becker-Döring difference equations can be written as[36,37]

$$\frac{\partial c_1}{\partial t} = -\sum_{n=1}^{n_m-1} a_n \left( c_1 c_n - \frac{\tilde{c}_1 \tilde{c}_n}{\tilde{c}_{n+1}} c_{n+1} \right), \quad (11)$$

$$\frac{\partial c_2}{\partial t} = \frac{1}{2} a_1 \left( c_1^2 - \frac{\tilde{c}_1^2}{\tilde{c}_2} c_2 \right) - a_2 \left( c_1 c_2 - \frac{\tilde{c}_1 \tilde{c}_2}{\tilde{c}_3} c_3 \right), \quad (12)$$

$$\frac{\partial c_n}{\partial t} = a_{n-1} \left( c_1 c_{n-1} - \frac{\tilde{c}_1 \tilde{c}_{n-1}}{\tilde{c}_n} c_n \right) - a_n \left( c_1 c_n - \frac{\tilde{c}_1 \tilde{c}_n}{\tilde{c}_{n+1}} c_{n+1} \right), \quad n = 3, \ldots n_m, \quad (13)$$

where tilde marks the quantities in the state of final equilibrium of the micellar solution.

For small deviations from equilibrium state of solution, the set of equations can be linearized by substitution $c_n(t) = \tilde{c}_n + \delta c_n(t)$, $\delta c_n(t)/\tilde{c}_n \ll 1$ for any $n \ge 1$. As has been shown in Ref.[36], it is convenient to rewrite the linearized Becker-Döring difference equations for aggregate concentrations in the form of the equation for time evolution of the vector $\mathbf{u}\ (u_1, u_2, \ldots, u_{n_m})$,

$$\partial \mathbf{u}/\partial t = \hat{\mathbf{A}} \mathbf{u}, \quad (14)$$

where $u_n(t) \equiv \delta c_n(t)/\sqrt{\tilde{c}_n}$, and the symmetric matrix $\hat{\mathbf{A}}$ has three-diagonal form with nonzero first row and column,

$$\hat{\mathbf{A}} \equiv \begin{pmatrix} \bullet & \bullet & \bullet & \bullet & \bullet & \cdots & \bullet & \bullet & \bullet \\ \bullet & \bullet & \bullet & 0 & 0 & \cdots & 0 & 0 & 0 \\ \bullet & \bullet & \bullet & \bullet & 0 & \cdots & 0 & 0 & 0 \\ \bullet & 0 & \bullet & \bullet & \bullet & \cdots & 0 & 0 & 0 \\ \bullet & 0 & 0 & \bullet & \bullet & \cdots & 0 & 0 & 0 \\ \vdots & \vdots & \vdots & \vdots & \vdots & \ddots & \vdots & \vdots & \vdots \\ \bullet & 0 & 0 & 0 & 0 & \cdots & \bullet & \bullet & 0 \\ \bullet & 0 & 0 & 0 & 0 & \cdots & \bullet & \bullet & \bullet \\ \bullet & 0 & 0 & 0 & 0 & \cdots & 0 & \bullet & \bullet \end{pmatrix} \quad (15)$$

where solid circles show distinct from zero elements determined as

$$A_{1,1} \equiv -a_1 \tilde{c}_1 - \sum_{n=1}^{n_m-1} a_n \tilde{c}_n, \quad (15a)$$

$$A_{1,k} = A_{k,1} \equiv \left( a_{k-1} \frac{\tilde{c}_{k-1}}{\tilde{c}_k} - a_k \right) \sqrt{\tilde{c}_1 \tilde{c}_k}, \quad k = 2 \ldots n_m, \quad (15b)$$

$$A_{k,k} \equiv -\left( a_{k-1} \frac{\tilde{c}_{k-1}}{\tilde{c}_k} + a_k \right) \tilde{c}_1, \quad k = 2 \ldots n_m, \quad (15c)$$

$$A_{k,k+1} = A_{k+1,k} \equiv a_k \tilde{c}_1 \sqrt{\frac{\tilde{c}_k}{\tilde{c}_{k+1}}}, \quad k = 2 \ldots n_m - 1. \quad (15d)$$

**Computed relaxation times and modes.** According to Eq.(15), all the eigenvalues of the matrix $\hat{\mathbf{A}}$ appear to be nonpositive and nondegenerated as in the cases of separate spherical or separate cylindrical micelles,[36,37] and we can rank their absolute values with the help of index $k$ as $\lambda_k$, $k = 0, 1, \ldots, n_m - 1$. Several consequent lowest absolute values of the eigenvalues computed with the help of eqs.(2)-(5), (10) and (15) for aggregation work 1 are shown in Fig.3. The final surfactant monomer concentration was taken here as $\tilde{c}_1 = (c_1)_{cmc_2}^{(w1)} = 0.9977$.

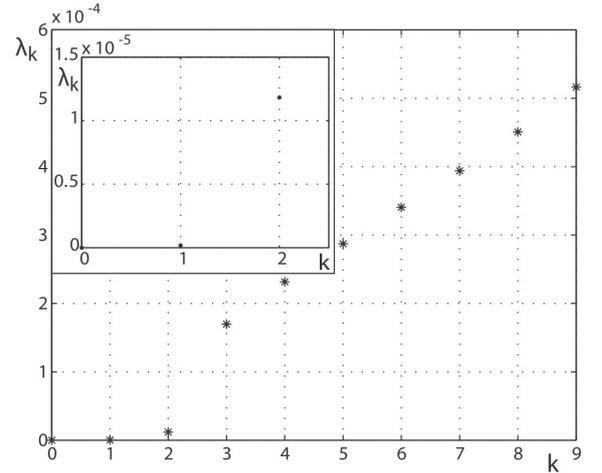

**Fig.3** The smallest moduli of eigenvalues of the matrix $\hat{\mathbf{A}}$ at relaxation to equilibrium (for aggregation work 1 and $\tilde{c}_1 = (c_1)_{cmc_2}^{(w1)} = 0.9977$).

As is seen from Fig.3, the smallest $\lambda_k$ equals zero and is marked by $k = 0$. It corresponds to the final equilibrium of the isolated micellar solution. The next two values $\lambda_1$ and $\lambda_2$ are very small but nonzero and corresponds to characteristic relaxation times $t_{s1} \equiv 1/\lambda_1$ and $t_{s2} \equiv 1/\lambda_2$. Evidently, these times are the largest among other relaxation times and thus can be called the times of slow relaxation. Since the values $\lambda_1$ and $\lambda_2$ differ more than by order of magnitude ($\lambda_1 \ll \lambda_2$ and $t_{s1} \gg t_{s2}$), there are two clearly distinct stages during the slow approach to final equilibrium. These stages can be called the first (with characteristic time $t_{s1}$) and second (with characteristic time $t_{s2}$) stages of slow relaxation. Existence of two stages of slow relaxation is a new specific feature of the system with coexisting spherical and cylindrical micelles.

The values $\lambda_k$ with indices $k = 3, 4, ...$ correspond to the times $1/\lambda_k$ of relaxation to an intermediate quasi-equilibrium of cylindrical micelles at $n > \bar{n}_0$ (to the right of the potential hill of the aggregation work in Fig.1). This intermediate equilibrium establishes much faster than slow relaxation proceeds, and this stage of relaxation can be called fast relaxation.

As follows from Fig.3, the time $t_f \equiv 1/\lambda_3$ is the largest among other specific times of fast relaxation, and can be called the time $t_f$ of fast relaxation. Other specific times $1/\lambda_k$ with $k \geq 4$ characterize more quick processes at fast relaxation.

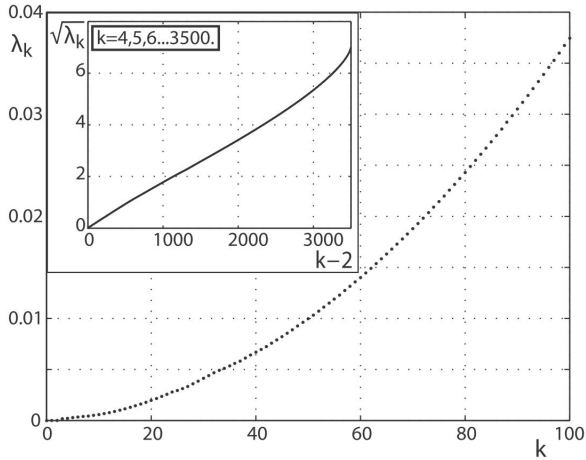

**Fig.4** Absolute values of eigenvalues of the matrix $\hat{\mathbf{A}}$ at relaxation to equilibrium at large numbers $k$ (for aggregation work 1 and $\tilde{c}_1 = (c_1)_{cmc_2}^{(w1)} = 0.9977$).

The values $\lambda_k$ with $k = 3, 4, ... 9$ are located in Fig.3 practically equidistantly. Thus for such numbers the dependence of the eigenvalues of the matrix $\hat{\mathbf{A}}$ on $k$ is almost linear. As follows from Fig.4, the linearity disturbs with growth of number $k$ to sufficiently large $k$, and the dependence of $\lambda_k$ on numbers $k$ can be considered for intermediate $k$ as quadratic.

An additional important information on relaxation in micellar solution can be obtained from the analysis of eigenvectors of matrix $\hat{\mathbf{A}}$. Eigenvectors $\mathbf{u}^{(k)}$, $k = 0, 1, ..., n_m - 1$, of matrix $\hat{\mathbf{A}}$ correspond to the relaxation modes $\delta c_n^{(k)}$ of nonequilibrium distribution of aggregates in aggregation number. These modes contribute to the overall nonequilibrium behavior of the deviation $\delta c_n(t) = \sum_{k=1} E_k \exp(-\lambda_k t) \delta c_n^{(k)}$ of concentrations $c_n(t)$ from equilibrium concentration $\tilde{c}_n$ at every aggregation number $n$ (coefficients $E_k$ are determined by initial conditions). For a discrete spectrum of $\lambda_k$, all $\delta c_n^{(k)}$ at different numbers $k$ are finite functions of aggregation number $n$.

Figures 5 show first five normalized relaxation modes $\delta c_n^{(k)}/\tilde{c}_n$, $k = 1, 2, 3, 4, 5$, computed as functions of the aggregation number $n$ for aggregation work 1 and monomer concentration $\tilde{c}_1 = (c_1)_{cmc_2}^{(w1)} = 0.9977$. As follows from Fig.5, modes with $k = 5, 4, 3$ describe the initial states for subsequent microstages of fast relaxation which develop in direction of smaller aggregation numbers, the smaller $k$ the shorter range in $n$ is occupied by non-monotonic variation of the corresponding mode $\delta c_n^{(k)}/\tilde{c}_n$. The mode $\delta c_n^{(5)}/\tilde{c}_n$ has 3 minima and 3 maxima and describes fast relaxation within the range $1 < n < 1500$, the mode $\delta c_n^{(4)}/\tilde{c}_n$ has 3 minima and 2 maxima and describes fast relaxation within the range $1 < n < 1200$ and so on. Fast relaxation completes with decay of the mode $\delta c_n^{(3)}$ and establishing three quasi-equilibrium states described by $\delta c_n^{(2)}$ in the ranges $1 < n < n_c^{(1)}$, $n_c^{(1)} < n < 180$ and $200 < n < 400$. The modes $\delta c_n^{(2)}$ and $\delta c_n^{(1)}$ decay with slow relaxation times $t_{s2}$ and $t_{s1}$.

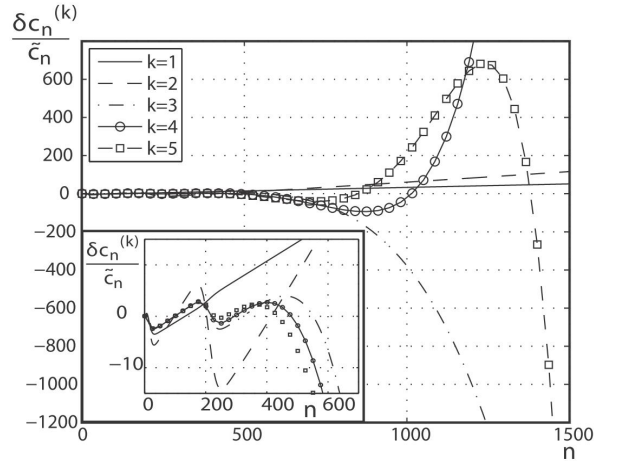

**Fig.5** Five normalized relaxation modes corresponding to the eigenvectors of matrix $\hat{\mathbf{A}}$ at $k = 1, 2, 3, 4, 5$ (for aggregation work 1 and $\tilde{c}_1 = (c_1)_{cmc_2}^{(w1)} = 0.9977$).

## IV. DEPENDENCE ON THE TOTAL SURFACTANT CONCENTRATION AND COMPARISON WITH PREDICTIONS OF ANALYTICAL KINETIC THEORY

Let us now consider the dependence of the

characteristic times and specific modes of relaxation on the total surfactant concentration. We will do this simultaniously comparing results with predictions of analytical kinetic theory based on the linearized differential Becker-Döring equation.

**Analytical kinetic theory.** The Becker-Döring difference equations (11)-(13) can be transformed in the continual limit at $n \gg 1$ to the differential equation

$$\frac{\partial c_n}{\partial t} = -\frac{\partial}{\partial n}\left\{a_n\tilde{c}_1\tilde{c}_n\left[\left(\frac{c_1}{\tilde{c}_1}-1\right)\frac{c_n}{\tilde{c}_n}-\frac{\partial}{\partial n}\frac{c_n}{\tilde{c}_n}\right]\right\}. \quad (16)$$

In order to find an analytical solution of eq.(16) at total surfactant concentrations above the cmc$_2$, it is convenient to consider separately the stages of slow and fast relaxation. As we mentioned yet, fast relaxation ends by establishing the separate quasi-equilibrium states for small premicellar aggregates with aggregation numbers $n < n_c^{(1)}$, for spherical micelles within range $n_c^{(1)} < n < n_c^{(2)}$ and for cylindrical aggregates with aggregation numbers $n > n_c^{(2)}$. It has been shown in the analytical theory[30,31,33] that the quasi-equilibrium states establish in fact in regions with aggregation numbers $n < n_c^{(1)} - \Delta n_c^{(1)}$, $n_c^{(1)} + \Delta n_c^{(1)} < n < n_c^{(2)} - \Delta n_c^{(2)}$ and $n > n_c^{(2)} + \Delta n_c^{(2)}$, where $\Delta n_c^{(1)}$ and $\Delta n_c^{(2)}$ are the half-widths of the first and second potential hill of the aggregation work $W_n$,

$$\left(\Delta n_c^{(1)}\right)^2 \equiv \frac{2}{\left|W_n''\right|_{n=n_c^{(1)}}}, \quad \left(\Delta n_c^{(2)}\right)^2 \equiv \frac{2}{\left|W_n''\right|_{n=n_c^{(2)}}}. \quad (17)$$

The difference of the local quasi-equilibrium states for premicellar aggregates, spherical and cylindrical micelles leads to the direct and backward fluxes of aggregates over first and second potential peaks of the aggregation work. At the stage of slow relaxation, these fluxes within the ranges $n_c^{(1)} - \Delta n_c^{(1)} < n < n_c^{(1)} + \Delta n_c^{(1)}$ and $n_c^{(2)} - \Delta n_c^{(2)} < n < n_c^{(2)} + \Delta n_c^{(2)}$ can be considered as quasi-steady.

**Slow relaxation.** By integration over all aggregation numbers and introducing the total concentration $c_{SM} = \int_{n_c^{(1)}}^{n_c^{(2)}} dn\, c_n$ of spherical micelles and the total concentration $c_{CM} = \int_{n_c^{(2)}}^{\infty} dn\, c_n$ of cylindrical micelles, eq.(16) can be converted to the balance equations

$$dc_{SM}/dt = J_1' - J_1'' - (J_2' - J_2''), \quad (18)$$

$$dc_{CM}/dt = J_2' - J_2'', \quad (19)$$

where quasi-steady direct fluxes $J_1'$ and $J_2'$ over the potential hills of the aggregation work $W_n$ (subscript referred to the potential peaks $W_c^{(1)}$ and $W_c^{(2)}$) and corresponding quasi-steady backward fluxes $J_1''$ and $J_2''$ over the same hills are defined[30,31,35]

$$J_1' = a_{n_c^{(1)}} c_1^2 \frac{\exp(-W_c^{(1)})}{\pi^{1/2}\Delta n_c^{(1)}}, \quad (20)$$

$$J_1'' = a_{n_c^{(1)}} c_1 c_{SM} \frac{\exp[-(W_c^{(1)} - W_s^{(1)})]}{\pi \Delta n_c^{(1)} \Delta n_s^{(1)}}, \quad (21)$$

$$J_2' = a_{n_c^{(2)}} c_1 c_{SM} \frac{\exp\left[-\left(W_c^{(2)} - W_s^{(1)}\right)\right]}{\pi \Delta n_c^{(2)} \Delta n_s^{(1)}}, \quad (22)$$

$$J_2'' = a_{n_c^{(2)}} c_1 c_{CM} \frac{\exp\left[-\left(W_c^{(2)} - W_0\right)\right]}{\pi^{1/2}\Delta n_c^{(2)}(n_* - \bar{n}_0)}, \quad (23)$$

where $\Delta n_s$ is the half-width of the first potential well of the aggregation work $W_n$,

$$\left(\Delta n_s^{(1)}\right)^2 \equiv \frac{2}{W_n''\big|_{n=n_s^{(1)}}}, \quad (24)$$

$n_* - \bar{n}_0$ is an analogue of the right half-width of the work $W_n$ at $n > \bar{n}_0$. As follows from eqs.(2) and (3) at $n > \bar{n}_0$, we can write

$$W_n = \bar{W}_0 + \bar{k}(n-\bar{n}_0) - (n-1)\ln c_1$$
$$= W_0 + \left(\bar{k} - \ln c_1\right)(n-\bar{n}_0) = W_0 + \frac{(n-\bar{n}_0)}{(n_* - \bar{n}_0)}, \quad (25)$$

thus

$$n_* - \bar{n}_0 = \left(\bar{k} - \ln c_1\right)^{-1}. \quad (26)$$

In a closed system, where equation $C = \sum_{n=1}^{n_m} n c_n$ holds, the total concentrations $c_{SM}$, $c_{CM}$ and surfactant monomer concentration $c_1$ are linked by the relation

$$C = c_1 + n_s c_{SM} + n_* c_{CM}. \quad (27)$$

For small deviations $\delta c_1(t)$, $\delta c_{SM}(t) \equiv c_{SM}(t) - \tilde{c}_{SM}$ and $\delta c_{CM}(t) \equiv c_{CM}(t) - \tilde{c}_{CM}$, eqs.(18), (19) and (26) can be reduced with the help of thermodynamic equalities

$$\frac{\partial W_n}{\partial c_1} = -\frac{n-1}{c_1}, \quad \frac{\partial n_s}{\partial c_1} = \frac{(\Delta n_s)^2}{2c_1}, \quad \frac{\partial n_*}{\partial c_1} = \frac{(n_* - \bar{n}_0)^2}{c_1} \quad (28)$$

to following coupled linear equations

$$\frac{d\delta c_{SM}}{dt} = -\alpha_{11}\delta c_{SM} - \alpha_{12}\delta c_{CM}, \quad (29a)$$

$$\frac{d\delta c_{CM}}{dt} = -\alpha_{21}\delta c_{SM} - \alpha_{22}\delta c_{CM}, \quad (29b)$$

where

$$\alpha_{11} \equiv \frac{\tilde{J}_1' + \tilde{J}_2'}{\tilde{c}_{SM}} + \frac{\tilde{n}_s^{(1)}\left(\tilde{n}_s^{(1)}\tilde{J}_1' - \left(\tilde{n}_* - \tilde{n}_s^{(1)}\right)\tilde{J}_2'\right)}{\tilde{c}_1 + \left(\Delta \tilde{n}_s^{(1)}\right)^2 \tilde{c}_{SM}/2 + (\tilde{n}_* - \bar{n}_0)^2 \tilde{c}_{CM}},$$

$$\alpha_{12} \equiv -\frac{\tilde{J}_2'}{\tilde{c}_{CM}} + \frac{\tilde{n}_*\left(\tilde{n}_s^{(1)}\tilde{J}_1' - \left(\tilde{n}_* - \tilde{n}_s^{(1)}\right)\tilde{J}_2'\right)}{\tilde{c}_1 + \left(\Delta \tilde{n}_s^{(1)}\right)^2 \tilde{c}_{SM}/2 + (\tilde{n}_* - \bar{n}_0)^2 \tilde{c}_{CM}},$$

$$\alpha_{21} \equiv -\frac{\tilde{J}_2'}{\tilde{c}_{SM}} + \frac{\tilde{n}_s^{(1)}\left(\tilde{n}_* - \tilde{n}_s^{(1)}\right)\tilde{J}_2'}{\tilde{c}_1 + \left(\Delta \tilde{n}_s^{(1)}\right)^2 \tilde{c}_{SM}/2 + (\tilde{n}_* - \bar{n}_0)^2 \tilde{c}_{CM}},$$

$$\alpha_{22} \equiv \frac{\tilde{J}'_2}{\tilde{c}_{CM}} + \frac{\tilde{n}_* \left(\tilde{n}_* - \tilde{n}_s^{(1)}\right)\tilde{J}'_2}{\tilde{c}_1 + \left(\Delta \tilde{n}_s^{(1)}\right)^2 \tilde{c}_{SM}/2 + \left(\tilde{n}_* - \bar{n}_0\right)^2 \tilde{c}_{CM}} \quad . \quad (30)$$

Solutions of eqs.(29) have the form

$$\delta c_{SM} = A_1 e^{-t/t_{s1}} + A_2 e^{-t/t_{s2}}, \quad (31a)$$

$$\delta c_{CM} = B_1 e^{-t/t_{s1}} + B_2 e^{-t/t_{s2}}. \quad (31b)$$

Here coefficients $A_1$, $A_2$, $B_1$, $B_2$ are determined by the initial conditions $\delta c_{SM}(0)$ and $\delta c_{CM}(0)$, the specific slow relaxation times $t_{s1}$ and $t_{s2}$ are expressed as

$$(t_{s1})^{-1} = \frac{1}{2}(\alpha_{11}+\alpha_{22}) - \left[\frac{1}{4}(\alpha_{11}-\alpha_{22})^2 + \alpha_{12}\alpha_{21}\right]^{1/2}, \quad (32a)$$

$$(t_{s2})^{-1} = \frac{1}{2}(\alpha_{11}+\alpha_{22}) + \left[\frac{1}{4}(\alpha_{11}-\alpha_{22})^2 + \alpha_{12}\alpha_{21}\right]^{1/2}. \quad (32b)$$

It should be noted that despite the the formulas (32) resemble the formulas (53) from ref.(47), their physical meaning is quite different because they describe coupling in relaxation between total concentration of spherical and cylindrical micelles.

Now we can compare the reciprocal specific times $t_{s1}^{-1}$ and $t_{s2}^{-1}$ with the moduli $\lambda_1$ and $\lambda_2$ of eigenvalues for matrix $\hat{A}$ computed in Section 2. Figures 6 and 7 show the dependence of $t_{s1}^{-1}$, $\lambda_1$ and $t_{s2}^{-1}$, $\lambda_2$ on equilibrium concentration $\tilde{c}_1$ of monomers for aggregation works 1 and 2 computed with the help of eqs.(32a), (32b), (30), (26), (20)-(24), (17), (10) and (3)-(7).

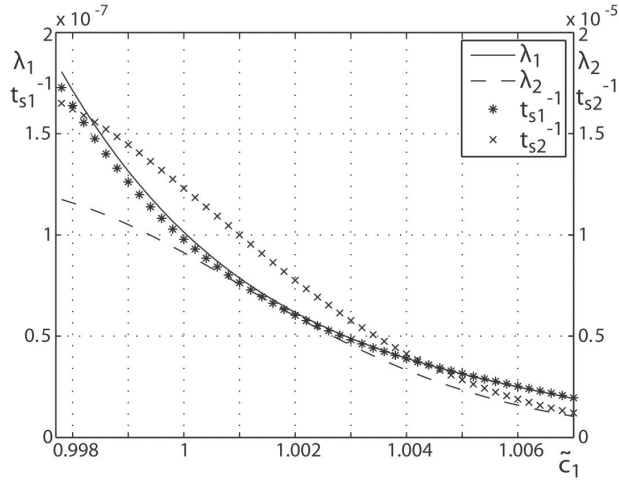

**Fig.6** Dependence of the reciprocal time $t_{s1}^{-1}$ and modulus $\lambda_1$ (their values are shown in the left vertical axis), the reciprocal time $t_{s2}^{-1}$ and modulus $\lambda_2$ (their values are shown in the right vertical axis) on equilibrium concentration $\tilde{c}_1$ of monomers for aggregation work 1.

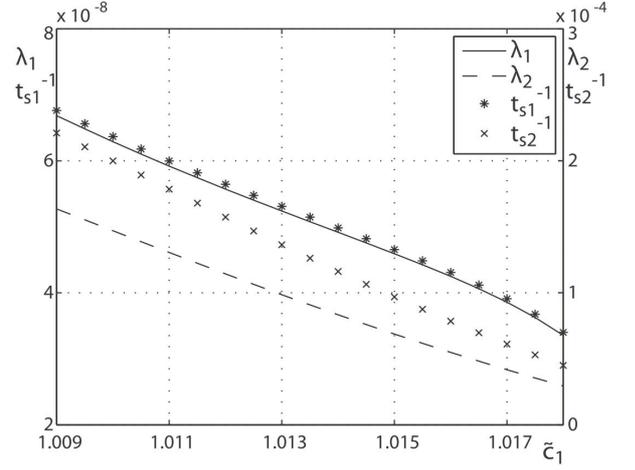

**Fig.7** Dependence of the reciprocal time $t_{s1}^{-1}$ and modulus $\lambda_1$ (their values are shown in the left vertical axis), the reciprocal time $t_{s2}^{-1}$ and modulus $\lambda_2$ (their values are shown in the right vertical axis) on equilibrium concentration $\tilde{c}_1$ of monomers for aggregation work 2.

As one can see from Figs.6 and 7, the analytical theory gives a fine prediction for both works in the case of $t_{s1}^{-1}$ and $\lambda_1$, even in the vicinity of cmc$_2$, where some assumptions of the theory weaken. The agreement of $t_{s2}^{-1}$ and $\lambda_2$ is worser in vicinity of cmc$_2$, but becomes better for larger total concentrations. The relative deviation of $t_{s2}^{-1}$ from $\lambda_2$ is smaller in the case of work 2. The time $t_{s1}$ is larger than the time $t_{s2}$ by two decimal orders. It means that corresponding stages in slow relaxation can be observable experimentally.

**Fast relaxation.** In terms of small relative deviations $\xi_n(t) \equiv \delta c_n(t)/\tilde{c}_n$, eq.(16) can be rewritten on the stage of fast relaxation in the linearized form as

$$\tilde{c}_n \frac{\partial \xi_n}{\partial t} = -\frac{\partial}{\partial n}\left[a_n \tilde{c}_1 \tilde{c}_n \left(\xi_1 - \frac{\partial \xi_n}{\partial n}\right)\right], \quad (33)$$

while the the condition (27) of fixed total concentration transforms to

$$\tilde{c}_1 \xi_1(t) = -\int_2^\infty dn\, n \tilde{c}_n \xi_n(t). \quad (34)$$

There were several models for the attachment coefficients $a_n$ which had been considered in the analysis of kinetic equations (33) for coexisting spherical and cylindrical micelles. For spherical micelles, i.e. at $n_c^{(1)} < n < n_c^{(2)}$, it was taken[33-35]

$$a_n = a_{n_s^{(1)}}, \quad (35)$$

and we will use the same approximation in eq.(33), assuming according to eq. (10) that

$$a_n = a_{\tilde{n}_s^{(1)}} = \left(\tilde{n}_s^{(1)}\right)^{1/3} (\tilde{n}_s^{(1)} + \bar{n}_0)^{2/3}/\bar{n}_0. \quad (36)$$

For cylindrical micelles, i.e. at $n \geq \bar{n}_0$, there were two approximations[33-35]

$$\begin{cases} a_n = a_{n_*} n/n_*, \\ a_n = a_{n_*}(n-\bar{n}_0)/(n_*-\bar{n}_0) \end{cases} \quad (37)$$

which suppose that total surfactant concentration is much larger than cmc$_2$ and $n_* \gg \bar{n}_0$. Here we will use in eq.(33) for $a_n$ at $n \geq \bar{n}_0$ a different approximation

$$a_n = (a_{\tilde{n}_*} - a_{\bar{n}_0})(n-\bar{n}_0)/(\tilde{n}_*-\bar{n}_0) + a_{\bar{n}_0} \quad (38)$$

which is more compatible with eq.(10) if $\tilde{n}_* - \bar{n}_0$, $a_{\tilde{n}_*}$ and $a_{\bar{n}_0}$ are calculated with the help of eqs. (26) and (10).

In the range $\tilde{n}_s^{(1)} - \Delta\tilde{n}_s^{(1)} < n < \tilde{n}_s^{(1)} + \Delta\tilde{n}_s^{(1)}$ of aggregation numbers for spherical micelles where, as follows from eq. (24), $W_n - \tilde{W}_s^{(1)} \simeq (n-\tilde{n}_s^{(1)})^2 / (\Delta\tilde{n}_s^{(1)})^2$, eq.(33) can be written in view of eqs.(1) and (36) in the dimensionless form as

$$\frac{(\Delta\tilde{n}_s^{(1)})^2}{a_{\tilde{n}_s^{(1)}}\tilde{c}_1}\frac{\partial\xi_n}{\partial t} = \frac{\partial^2\xi_n}{\partial r^2} - 2r\frac{\partial\xi_n}{\partial r} + 2\Delta\tilde{n}_s^{(1)}r\xi_1 \quad (39)$$

where

$$r \equiv (n-\tilde{n}_s^{(1)})/\Delta\tilde{n}_s^{(1)}. \quad (40)$$

Correspondingly, in the range $n > \bar{n}_0$ of aggregation numbers for cylindrical micelles where eq. (25) is valid, eq.(33) can be written in view of eqs.(1) and (38) in the dimensionless form as

$$\frac{(\tilde{n}_*-\bar{n}_0)^2}{(a_{\tilde{n}_*}-a_{\bar{n}_0})\tilde{c}_1}\frac{\partial\xi_n}{\partial t} = s\frac{\partial^2\xi_n}{\partial s^2} + (1-s)\frac{\partial\xi_n}{\partial s} - $$
$$-(\tilde{n}_*-\bar{n}_0)(1-s)\xi_1 + \frac{a_{\bar{n}_0}}{a_{\tilde{n}_*}-a_{\bar{n}_0}}\left[\frac{\partial^2\xi_n}{\partial s^2} - \frac{\partial\xi_n}{\partial s} + (\tilde{n}_*-\bar{n}_0)\xi_1\right]. \quad (41)$$

where

$$s \equiv (n-\bar{n}_0)/(\tilde{n}_*-\bar{n}_0). \quad (42)$$

According to eqs.(33) and (34), nonuniform differential equations (39) and (41) are coupled through relative deviation of concentration of surfactant monomers. Assuming that the main contribution to the right-hand side of eq.(34) is given by spherical and cylindrical micelles (neglecting the contributions of premicellar aggregages with $n < \tilde{n}_s^{(1)} - \Delta\tilde{n}_s^{(1)}$ and transient aggregates with $\tilde{n}_s^{(1)} + \Delta\tilde{n}_s^{(1)} < n < \bar{n}_0$), one can rewrite eq.(34) in view of eq.(1) as

$$\xi_1(t) = -e^{-\tilde{W}_s^{(1)}}\int_{-\tilde{n}_s^{(1)}-\Delta\tilde{n}_s^{(1)}}^{-\tilde{n}_s^{(1)}+\Delta\tilde{n}_s^{(1)}} dn\, n\, e^{-(n-\tilde{n}_s^{(1)})^2/(\Delta\tilde{n}_s^{(1)})^2}\xi_n(t) - $$
$$-e^{-\tilde{W}_0}\int_{\bar{n}_0}^{\infty} dn\, n\, e^{-(n-\bar{n}_0)/(\tilde{n}_*-\bar{n}_0)}\xi_n(t). \quad (43)$$

Solution of eq.(39) can be represented as series in the Hermite, $H_i(r)$ $(i=1,2,...)$, orthogonal polynomials. As follows from eq. (41), its solution can be represented as series in the Laguerre, $L_i(s)$ $(i=1,2,...)$, orthogonal polynomials only in the case when

$$\frac{a_{\bar{n}_0}}{a_{\tilde{n}_*}-a_{\bar{n}_0}} \ll 1, \quad (44)$$

and we can neglect the corresponding last term on the right-hand side of eq.(41). Then the solution of eqs.(39) and (41) can be written as

$$\xi_n(t) = \begin{cases} \sum_{i=1}^{\infty} m_i(t) H_i\left(\frac{n-\tilde{n}_s^{(1)}}{\Delta\tilde{n}_s^{(1)}}\right) & (|n-\tilde{n}_s^{(1)}| \leq \Delta\tilde{n}_s^{(1)}), \\ \sum_{i=1}^{\infty} q_i(t) L_i\left(\frac{n-\bar{n}_0}{\tilde{n}_*-\bar{n}_0}\right) & (n \geq \bar{n}_0). \end{cases} \quad (45)$$

Because the total concentrations of spherical and cylindrical micelles are preserved at the stage of fast relaxation, it can be easily proved[33-35] that coefficients $m_0$ and $q_0$ in eq.(45) are absent. Substitution of eq.(45) into eq.(43) gives the following expression for deviation $\xi_1(t)$ of surfactant monomer concentration

$$\xi_1(t) = -e^{-\tilde{W}_s^{(1)}}\sqrt{\pi}(\Delta\tilde{n}_s^{(1)})^2 m_1(t) + e^{-\tilde{W}_0}(\tilde{n}_*-\bar{n}_0)^2 q_1(t). \quad (46)$$

Substituting eqs.(45) and (46) in eqs.(39) and (41), taking into account strong inequality (44) and using the orthogonality of the Hermite and Laguerre polynomials lead to the following differential equations for time-dependent coefficients $m_i(t)$ and $q_i(t)$:

$$\frac{dm_1(t)}{dt} = -\beta_{11}m_1(t) - \beta_{12}q_1(t), \quad (47a)$$

$$\frac{dq_1(t)}{dt} = -\beta_{21}m_1(t) - \beta_{22}q_1(t), \quad (47b)$$

$$\frac{dm_i(t)}{dt} = -\frac{2ia_{\tilde{n}_s^{(1)}}\tilde{c}_1}{(\Delta\tilde{n}_s^{(1)})^2}m_i(t) \quad (i=2,3,...), \quad (48a)$$

$$\frac{dq_i(t)}{dt} = -\frac{i(a_{\tilde{n}_*}-a_{\bar{n}_0})\tilde{c}_1}{(\tilde{n}_*-\bar{n}_0)^2}q_i(t) \quad (i=2,3,...). \quad (48b)$$

where

$$\beta_{11} \equiv \frac{2a_{\tilde{n}_s^{(1)}}\tilde{c}_1}{(\Delta\tilde{n}_s^{(1)})^2}\left[1+e^{-\tilde{W}_s^{(1)}}\frac{\sqrt{\pi}}{2}(\Delta\tilde{n}_s^{(1)})^3\right],$$

$$\beta_{12} \equiv -\frac{a_{\tilde{n}_s^{(1)}}\tilde{c}_1}{\Delta\tilde{n}_s^{(1)}}e^{-\tilde{W}_0}(\tilde{n}_*-\bar{n}_0)^2,$$

$$\beta_{21} \equiv -\frac{(a_{\tilde{n}_*}-a_{\bar{n}_0})\tilde{c}_1}{(\tilde{n}_*-\bar{n}_0)}e^{-\tilde{W}_s^{(1)}}\sqrt{\pi}(\Delta\tilde{n}_s^{(1)})^2,$$

$$\beta_{22} \equiv \frac{(a_{\tilde{n}_*}-a_{\bar{n}_0})\tilde{c}_1}{(\tilde{n}_*-\bar{n}_0)^2}\left[1+e^{-\tilde{W}_0}(\tilde{n}_*-\bar{n}_0)^3\right]. \quad (49)$$

Integration of eqs.(47) and (48) gives

$$m_1(t) = D_1 e^{-t/\tau_{s1}} + D_2 e^{-t/\tau_{c1}}, \quad (50a)$$

$$q_1(t) = F_1 e^{-t/\tau_{s1}} + F_2 e^{-t/\tau_{c1}}, \quad (50b)$$

$$m_i(t) = m_i(0) e^{-t/\tau_{si}}, \quad q_i(t) = q_i(0) e^{-t/\tau_{ci}} \quad (i=2,3,...) \quad (51)$$

where

$$\tau_{s1}^{-1} = \frac{1}{2}(\beta_{11} + \beta_{22}) + \left[\frac{1}{4}(\beta_{11} - \beta_{22})^2 + \beta_{12}\beta_{21}\right]^{1/2}, \quad (52a)$$

$$\tau_{c1}^{-1} = \frac{1}{2}(\beta_{11} + \beta_{22}) - \left[\frac{1}{4}(\beta_{11} - \beta_{22})^2 + \beta_{12}\beta_{21}\right]^{1/2}, \quad (52b)$$

$$\tau_{si} = \frac{\left(\Delta \tilde{n}_s^{(1)}\right)^2}{2 i a_{\tilde{n}_s^{(1)}} \tilde{c}_1}, \quad \tau_{ci} = \frac{(\tilde{n}_* - \bar{n}_0)^2}{i(a_{\tilde{n}_*} - a_{\bar{n}_0}) \tilde{c}_1} \quad (i=2,3,...), \quad (53)$$

coefficients $D_1$, $D_2$, $F_1$ and $F_2$ are determined by the initial conditions $m_1(0)$ and $q_1(0)$.

Dependences of the analytical reciprocal specific times of fast relaxation $\tau_{ci}^{-1}$ and $\tau_{si}^{-1}$ $(i=2,3,..7)$ and moduli $\lambda_k$ $(k=3,4,...)$ corresponding to the eigenvalues of the matrix $\hat{\mathbf{A}}$ on equilibrium monomer concentration $\tilde{c}_1$ are shown in Figs.8 and 9. We have used for computation eqs.(52), (53), (49), (26), (24), (10) and models (3)-(7) for aggregation works 1 and 2.

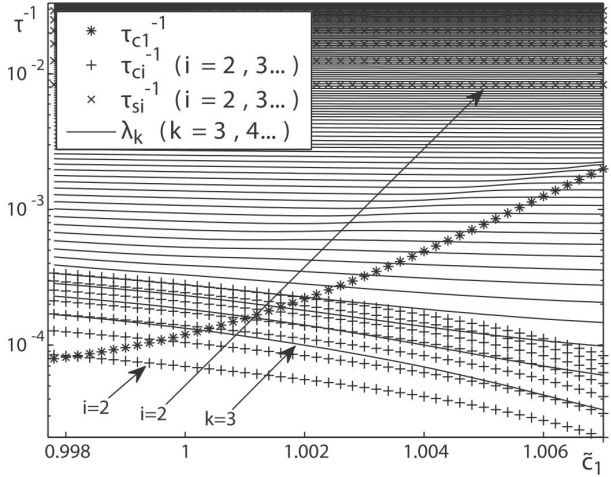

**Fig.8** The dependence of the analytical reciprocal specific times of fast relaxation $\tau_{ci}^{-1}$ and $\tau_{si}^{-1}$ $(i=2,3,..7)$ and moduli $\lambda_k$ $(k=3,4,...)$ on equilibrium concentration $\tilde{c}_1$ of monomers for aggregation work 1. Indices increase from below to top.

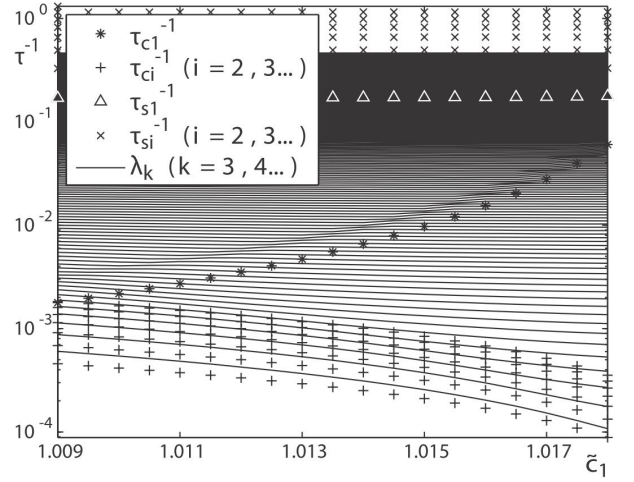

**Fig.9** The dependence of the analytical reciprocal specific times of fast relaxation $\tau_{ci}^{-1}$ and $\tau_{si}^{-1}$ $(i=2,3,..7)$ and moduli $\lambda_k$ $(k=3,4,...)$ on equilibrium concentration $\tilde{c}_1$ of monomers for aggregation work 2. Indices increase from below to top.

As follows from behaviour of the curves shown in Figs.8 and 9, the relation of discrete and analytical computations requires a special comment. First of all, inequality (44), which was used for simplification of eq.(41) and expressing the concentration of cylindrical micelles as the series in the Laguerre orthogonal polynomials, is strictly satisfied at $\tilde{n}_* \gg \bar{n}_0$ only. Moreover, the region of nonlinear behaviour of work $W_n$ between aggregation numbers $\tilde{n}_c^{(2)}$ and $\bar{n}_0$ has been excluded from analytical consideration. For the parameters (4), (5) of the model of the aggregation work 1 with $\bar{n}_0 = 301$, the aggregation number $\tilde{n}_*$ changes from 400 to 764 within the range of equilibrium monomer concentration $\tilde{c}_1$ in Fig.8, and the total number of micelles in the region between aggregation numbers $\tilde{n}_c^{(2)}$ and $\bar{n}_0$ is comparable to the total number of cylindrical micelles. For the parameters (6),(7) of the model of the aggregation work 2 with $\bar{n}_0 = 50$, the aggregation number $\tilde{n}_*$ changes from 150 to 513 within the range of equilibrium monomer concentration $\tilde{c}_1$ in Fig.9, and the total number of micelles in the region between aggregation numbers $\tilde{n}_c^{(2)}$ and $\bar{n}_0$ is smaller than the total number of cylindrical micelles. With increasing the total surfactant concentration, the ratio $\tilde{n}_*/\bar{n}_0$ increases and the relative share of the micelles in the region $[\tilde{n}_c^{(2)}, \bar{n}_0]$ decreases. These facts explain why an agreement between smallest analytical reciprocal times $\tau_{ci}^{-1}$ $(i=2,3,4)$ of fast relaxation and moduli $\lambda_k$ for $k=3,4,5$ is more visible in Fig.9 than in Fig.8 and improves with growth of the surfactant concentration. It should be noted here that we can expect to find an agreement only for lowest reciprocal times and moduli of computed eigenvalues, because we used in analytical treatment a continual approach with boundaries extended formally to infinity and simlified parabolic and linear approximations for the aggregation work. As follows from Figs.8 and 9, the largest specific time of fast relaxation is determined in analytical theory as $\tau_{c2}$ which corresponds to reciprocal modulus $\lambda_3^{-1}$ of computed eigenvalues of the matrix (15).

**Switching between fast relaxation times and modes.** With growth of concentration $\tilde{c}_1$ in Fig.8 and 9, the curve for $\tau_{c1}^{-1}$ intersects curves for $\tau_{ci}^{-1}$ ($i > 2$ in Fig.8 and $i > 8$ in Fig.9) and $\lambda_k$ ($k > 2$ in Fig.8 and $k > 8$ in Fig.9). It is also seen from Fig.8 and 9, that the curves for lowest moduli of eigenvalues of matrix (15) are monotonic functions of equilibrium monomer concentration, but with increasing the number $k$, these curves become non-monotonic with inflexion point. Upper part of the solid curves for eigenvalues $\lambda_k$ after the inflection point looks to be continued as a lower part of the solid curves for $\lambda_{k+1}$. One may expect that such switching in the eigenvalues of matrix $\hat{\mathbf{A}}$ should be accompanied by corresponding transition in the relaxation modes. As follows from Figs.8 and 9, the curve for $\tau_{c1}^{-1}$ appears to be close to the envelope for the inflection points of transitions between lower and upper parts of moduli $\lambda_k$ of the discrete matrix $\hat{\mathbf{A}}$ with larger number $k$. But for smaller equilibrium monomer concentrations, the curve for $\tau_{c1}^{-1}$ can be smaller than $\tau_{c2}^{-1}$. Does this mean that there is a switching in the slowest fast relaxation mode at some total surfactant concentration even for monotonically varying eigenvalue of the discrete matrix $\hat{\mathbf{A}}$ ?

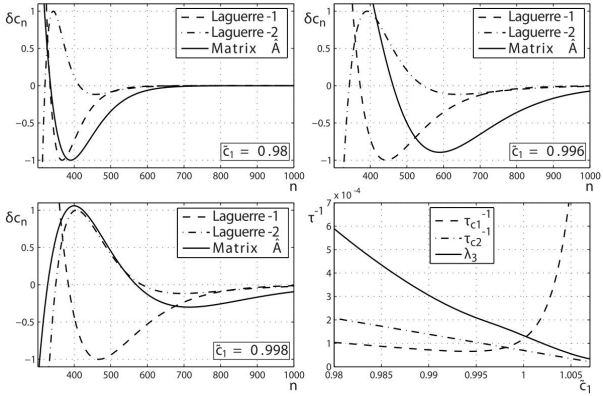

**Fig.10** Switching in relaxation mode $\delta c_n^{(3)}$ and behavior of analytical reciprocal times $\tau_{c1}^{-1}, \tau_{c2}^{-1}$ and computed modulus $\lambda_3$ with increasing final equilibrium monomer concentration $\tilde{c}_1$ for work 1.

Figure 10 gives an answer to this question. Solid lines in Fig.10 show variation of the dependence of the mode $\delta c_n^{(3)}$ (computed for discrete matrix (15) with using eqs.(3)-(5) and (10) for model of aggregation work 1) on the aggregation number $n$ with increase in equilibrium monomer concentration $\tilde{c}_1$ value from 0.98 to 0.998. The dash-dotted and dashed lines in Fig.10 depict the behavior of expressions at $n \geq \bar{n}_0$ for the analytical modes $L_1\left(\frac{n-\bar{n}_0}{\tilde{n}_* - \bar{n}_0}\right)\tilde{c}_n$ and $L_2\left(\frac{n-\bar{n}_0}{\tilde{n}_* - \bar{n}_0}\right)\tilde{c}_n$, respectively, which give according to eq.(45) a contribution to $\delta c_n = \xi_n \tilde{c}_n$ from cylindrical micelles. We need not consider here the contribution of spherical micelles within $\left|n - \tilde{n}_s^{(1)}\right| \leq \Delta \tilde{n}_s^{(1)}$, because, as follows from eqs.(52) and (53), fast relaxation of spherical micelles becomes shorter than relaxation of cylindrical micelles if $\tau_{c1}^{-1} > \tau_{c2}^{-1}$. As we can see from Fig.10, the behavior of the mode $\delta c_n^{(2)}$ at $\tilde{c}_1 = 0.98$ is close to that of analytical $L_1\left(\frac{n-\bar{n}_0}{\tilde{n}_* - \bar{n}_0}\right)\tilde{c}_n$. With increase in $\tilde{c}_1$, the curve of the mode $\delta c_n^{(2)}$ shifts towards curve of analytical $L_2\left(\frac{n-\bar{n}_0}{\tilde{n}_* - \bar{n}_0}\right)\tilde{c}_n$ and becomes very close to this curve starting from $\tilde{c}_1 = 0.998$. Such switching of the mode $\delta c_n^{(2)}$ is confirmed by last slide in Fig.10 showing the behavior of analytical reciprocal times $\tau_{c1}^{-1}$, $\tau_{c2}^{-1}$ and computed modulus $\lambda_3$ as functions of equilibrium monomer concentration. As follows from this slide, $\tau_{c1}^{-1} < \tau_{c2}^{-1}$ for $\tilde{c}_1$ value from 0.98 to 0.998 and $\tau_{c1}^{-1} > \tau_{c2}^{-1}$ for $\tilde{c}_1$ value above 0.998. As a consequence, the analytical mode $q_1 L_1\left(\frac{n-\bar{n}_0}{\tilde{n}_* - \bar{n}_0}\right)\tilde{c}_n$ should be the slowest mode in fast relaxation for $\tilde{c}_1$ value from 0.98 to 0.998, while the analytical mode $q_2 L_2\left(\frac{n-\bar{n}_0}{\tilde{n}_* - \bar{n}_0}\right)\tilde{c}_n$ becomes the slowest at $\tilde{c}_1$ value above 0.998. Just such behavior has been illustrated by first three slides in Fig.10. Finally we can say that analytical representation of the modulus $\lambda_3$ consists of two pieces ($\tau_{c1}^{-1}$ at $0.98 < \tilde{c}_1 < 0.998$ and $\tau_{c2}^{-1}$ at $\tilde{c}_1 > 0.998$) and the analytical desription of the slowest mode of fast relaxation for coexisting spherical and cylindrical micelles is satisfactory even in the vicinity of cmc$_2$.

## V. CONCLUSIONS

This study shows that eigenvalues and eigenvectors of the matrix of coefficients in the linearized discrete Becker-Döring kinetic equations applied to aggregation in surfactant solution give the full picture of multi-scale micellar relaxation in solutions with coexisting spherical and cylindrical micelles. We employed here the Smoluchowsky stationary diffusion model for the attachment rates of surfactant monomers to surfactant aggregates with matching the rates for spherical aggregates and the rates for large cylindrical micelles and some rather simple interpolation formula for the aggregation work as a function of aggregation number, but our approach can be extended to more realistic models if they will be proposed, including the case of fusion–fission of aggregates and using the generalized discrete Smoluchowsky kinetic equation.[29] The dependence of relaxation times and modes on the total surfactant concentration has been found for total surfactant concentrations within the wide range from the vicinity of the cmc$_2$ and well above the cmc$_2$. A special analysis has been done for coupling between nonequilibrium concentrations of spherical and cylindrical aggregates at slow and fast relaxation and specific additional relaxation times and modes which appear due to this coupling. The results of computations have been compared with the results of the analytical theory for coexisting spherical and cylindrical micelles obtained recently under some limitations for the continuous Becker-Döring kinetic equation. These results demonstrated a fair agreement even in the vicinity of the cmc$_2$ where the analytical theory looses formally its applicability.

## ACKNOWLEDGMENTS

This work was supported by St. Petersburg State University (grant 11.37.183.2014), grant RFBR 13-03-00991a, and the program of Russian Academy of Sciences "Chemistry and Physico-Chemistry of Supramolecular Systems and Atomic Clusters".

* Author to whom correspondence should be addressed. Electronic mail: akshch@list.ru.

[1] M.J. Rosen and J.T. Kunjappu, *Surfactants and Interfacial Phenomena*, 4th ed. (John Wiley & Sons, Inc.: Hoboken, NJ, 2012).
[2] K.L. Mittal and D.O. Shah, *Adsorption and aggregation of surfactants in solution* (Marcel Dekker, Inc.: New York, 2003).
[3] K. Holmberg, B. Jonsson, B. Kronberg and B. Lindman, *Surfactants and Polymers in Aqueous Solution*, 2nd ed. (John Wiley & Sons, Ltd: New York, 2002).
[4] A.K. Shchekin, F.M. Kuni, A.P. Grinin and A.I. Rusanov, Nucleation in micellization processes. In *Nucleation Theory and Applications*, edited by J. W. P. Schmelzer (Wiley: New York, 2005), Chap.9, p.312.
[5] R. Hadgiivanova and H. Diamant, J. Chem. Phys. **130**, 114901 (2009).
[6] R. Hadgiivanova, H. Diamant and D. Andelman, J. Phys. Chem. B **115**, 7268 (2011).
[7] R. Zana, Dynamics in Micellar Solutions of Surfactants. In *Dynamics of Surfactant Self-Assembles, Micelles, Microemulsions, Vesicles, and Lyotrophic Phases,* Surfactant Science Series Vol.125, edited by R. Zana (Taylor & Francis, Boca Raton, 2005), Chap 3, p.75.
[8] M. Gradzielski, Curr. Opin. Colloid Interface Sci. **8**, 337 (2003).
[9] K.D. Danov, P.A. Kralchevsky, N.D. Denkov and K.P. Ananthapadmanabhan, Adv. Colloid Interface Sci. **119**, 1 (2006).
[10] E.A.G. Aniansson and S.N. Wall, J. Phys. Chem. **78**, 1024 (1974).
[11] E.A.G. Aniansson and S.N. Wall, J. Phys. Chem. **79**, 857 (1975).
[12] S.N. Wall, and E.A.G. Aniansson, J. Phys. Chem. **84**, 727 (1980).
[13] M. Almgren, E.A.G. Aniansson and K. Holmaker, Chem. Phys. **19**, 1 (1977).
[14] M. Teubner, J. Phys. Chem. **83**, 2917 (1979).
[15] M. Kahlweit, Pure Appl. Chem. **53**, 2069 (1981).
[16] M. Kahlweit and M. Teubner, Adv. Colloid Interface Sci. **13**, 1 (1980).
[17] R. Becker and W. Döring, Ann. Phys. **24**, 719 (1935).
[18] R. Becker, *Theory of Heat*; 2nd ed. (Springer-Verlag: Berlin-Heidelberg-New York, 1967).
[19] J.M. Ball, J. Carr and O. Penrose, Commun. Math. Phys. **104**, 657 (1986).
[20] F.M. Kuni, A.I. Rusanov, A.K. Shchekin and A.P. Grinin, Russ. J. Phys. Chem. **79**, 833 (2005).
[21] I.M. Griffiths, C.D. Bain, C.J.W. Breward, D.M. Colegate, P.D. Howell and S.L. Waters, J. Colloid and Interface Sci. **360**, 662 (2011).
[22] I.M. Griffiths, C.D. Bain, C.J.W. Breward, S.J. Chapman, P.D. Howell and S.L. Waters, SIAM Journal on Applied Mathematics **72**, 201 (2012).
[23] I.M. Griffiths, C.J.W. Breward, D.M. Colegate, P.J. Dellar, P.D. Howella and C.D. Bain, Soft Matter **9**, 853 (2013).
[24] G.V. Jensen, R. Lund, J. Gummel, M. Monkenbusch, T. Narayanan and J.S. Pedersen, J. Am. Chem. Soc. **135**, 7214 (2013).
[25] R. Lund, L. Willner, M. Monkenbusch, P. Panine, T. Narayanan, J. Colmenero and D. Richter, Phys. Rev. Lett. **102**, 188301 (2009).
[26] G. Waton, J. Phys. Chem. B **101**, 9727 (1997).
[27] R. Pool and P.G. Bolhuis, Phys. Rev. Lett. **97**, 018302 (2006).
[28] R. Pool, P.G. Bolhuis, J. Chem. Phys. **126**, 244703 (2007).
[29] A.K. Shchekin, M.S. Kshevetskiy and O.S. Pelevina, Colloid J. **73**, 406 (2011).
[30] F.M. Kuni, A.K. Shchekin, A.P. Grinin and A.I. Rusanov, Colloid J. **67**, 41 (2005).
[31] F.M. Kuni, A.K. Shchekin, A.I. Rusanov and A.P. Grinin, Langmuir **22**, 1534 (2006).
[32] M.S. Kshevetskii, A.K. Shchekin and F.M. Kuni, Colloid J. **70**, 455 (2008).
[33] A.K. Shchekin, F.M. Kuni, A.P. Grinin and A.I. Rusanov, Russ. J. Phys. Chem. A **82**, 101 (2008).
[34] M.S. Kshevetskiy and A.K. Shchekin, J. Chem. Phys. **131**, 074114 (2009).
[35] A.K. Shchekin, A.I. Rusanov and F.M. Kuni, Chemistry Letters **41**, 1081 (2012).
[36] I.A. Babintsev, L.Ts. Adzhemyan and A.K. Shchekin, J. Chem. Phys. **137**, 044902 (2012).
[37] I.A. Babintsev, L.Ts. Adzhemyan and A.K. Shchekin, Soft Matter **10**, 2619 (2014).
[38] C. Tanford, *The Hydrophobic Effect*, 2nd ed. (Wiley: New York, 1980).
[39] R. Nagarajan and E. Ruckenstein, Self-Assembled Systems. In *Equations of State for Fluids and Fluid Mixtures*; J.V. Sengers, R.F. Kayser, C.J. Peters, H.J. White, Jr., Eds. Experimental Thermodynamics, Vol.5; (Elsevier Science: Amsterdam, 2000), Chap 15, p.589.
[40] R. Nagarajan, Theory of Micelle Formation: Quantitative Approach to Predicting Micellar Properties from Surfactant Molecular Structure. In *Structure-Performance Relationships in Surfactants*, K. Esumi and M. Ueno, Eds. Surfactant Science Series Vol. 112 (Marcel Dekker, 2003), Chap 1, p.1.
[41] A.I.Rusanov, F.M. Kuni, A.P. Grinin and A.K. Shchekin, Colloid J. **64**, 605 (2002).
[42] G. Porte, Y. Poggi, J. Appell and G. Maret, J. Phys. Chem. **88**, 5713 (1984).
[43] S. May and A. Ben-Shaul, J. Phys. Chem. B **105**, 630 (2001).
[44] S. May and A. Ben-Shaul, Molecular Packing in Cylindrical Micelles. In: *Giant Micelles: Properties and Applications* R. Zana and E. Kaler, Eds. (CRC Press: Boca Raton, 2007), p.41.
[45] M. Smoluchowski, Phys. Chem. **92**, 129 (1917).
[46] G. Mohan and D.I. Kopelevich, J. Chem. Phys. **128**, 044905 (2008).
[47] D.G. Hall, J. Chem Soc., Faraday Trans. 1 **83**, 967 (1987).